\documentclass[aps,prb,superscriptaddress,floatfix,reprint]{revtex4-1}
\usepackage{epsfig}
\usepackage{amsmath}
\usepackage{amsfonts}
\usepackage{amssymb}
\usepackage{mathptmx}
\usepackage{eucal}
\usepackage{bm}

\DeclareMathOperator{\const}{\mathop{\mathrm{const}}}
\newcommand{\Eq}[1]{Eq.\ (\ref{#1})}
\newcommand{\Eqs}[1]{Eqs.\ (\ref{#1})}
\begin{document}

\title{Resonant subgap current transport in Josephson field effect transistor}

\author{E.~V.~Bezuglyi}
\affiliation{B.Verkin Institute for Low Temperature Physics and Engineering,
Kharkiv 61103, Ukraine}
\affiliation{Chalmers University of Technology, S-41296 G\"oteborg, Sweden}
\author{E.~N.~Bratus'}
\affiliation{B.Verkin Institute for Low Temperature Physics and Engineering,
Kharkiv 61103, Ukraine}
\author{V.~S.~Shumeiko}
\affiliation{Chalmers University of Technology, S-41296
G\"oteborg, Sweden}

\begin{abstract}
We study theoretically the current-voltage characteristics (IVCs) of the Josephson field effect transistor --- a ballistic SNINS junction with superconducting (S) electrodes confining a planar normal metal region (N), which is controlled by the gate induced potential barrier (I). Using the computation technique developed earlier for long single-channel junctions in the coherent multiple Andreev reflections (MAR) regime, we find significant difference of the subgap current structure compared to the subharmonic gap structure in tunnel junctions and atomic-size point contacts. For long junctions, whose length significantly exceeds the coherence length, the IVC exhibits current peaks at multiples (harmonics) of the distance $\delta_m$ between the static Andreev levels, $eV_n = n\delta_m$. Moreover, the averaged IVC follows the power-like behavior rather than the exponential one, and has a universal scaling with the junction transparency. This result is qualitatively understood using an analytical approach based on the concept of resonant MAR trajectories. In shorter junctions having length comparable to the coherence length, the IVC has an exponential form common for point contacts, however the current structures appear at the subharmonics of the interlevel distance, $eV_n = \delta_m/n$, rather than the gap subharmonics $2\Delta/n$.
\end{abstract}

\maketitle

\section{Introduction}
In a number of experiments, the current-voltage characteristics (IVCs) were
measured in superconductor-normal metal-superconductor (SNS) junctions
consisting of high-mobility two-dimensional electron gas (2DEG) connected to
two bulk superconducting electrodes.\cite{Taka1,Taka,Schapers,Chrestin,Taka2,Hamburg} In these devices, the electrodes have a large width $w$ up to $40$ $\mu$m, while the distance $L$ between the electrodes varies between $0.2$--$1$ $\mu$m. This is comparable or larger than the coherence length $\xi_N=\hbar v_F/\Delta$, where $v_F$ is the Fermi velocity in the 2DEG, but much smaller than the elastic and inelastic scattering lengths. These junctions show well pronounced Josephson effect and have rather small NS interface resistance which corresponds to a large transmission coefficient $\sim 0.8$.\cite{Chrestin,Hamburg}
\begin{figure}[b]\vspace{-4mm}
\centerline{\epsfxsize=8cm\epsffile{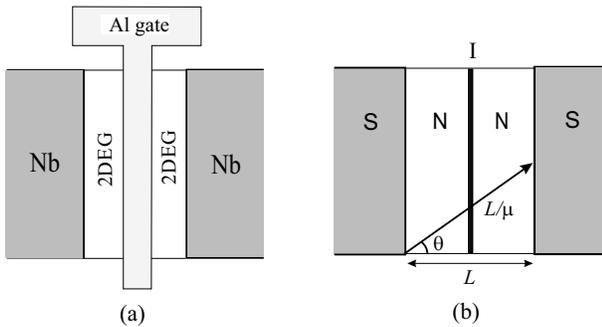}}\vspace{-4mm}
\caption{Experimental JOFET setup\cite{Taka2} (a) and the theoretical model (b) of the gated SNS junction, the role of a controlled tunnel barrier (I) is played by the gate potential. The tilted arrow depicts electron trajectory having length $L/\mu$, $\mu = \cos\theta$.}
\label{jofet}
\end{figure}

Most of these 2DEG structures were fabricated in effort to develop a Josephson field effect transistor (JOFET).\cite{Clark} In this device, schematically shown in Fig.~\ref{jofet}(a) (see, e.g., Ref.~\onlinecite{Taka2}), the Josephson current is controlled by the gate voltage which changes the carrier concentration within the area just beneath the gate, thus creating an effective potential barrier across the 2DEG layer. Such device is a practical realization of the theoretical model of a wide SNINS junction with tunable potential barrier inside the 2D normal metal layer sketched in Fig.~\ref{jofet}(b), provided the barrier width is much smaller than $L$. The dc and ac Josephson
currents in such junctions were studied in Refs.~\onlinecite{Zaikin,Zaikin2,Zaikin3}. In our paper, we examine the dissipative charge transport through JOFET in the regime of given applied voltage and predict a markedly different subgap current structure compared to point contacts and tunnel junctions, originating from Andreev bound states in the normal layer.

Assuming the dephasing length much larger than the junction length, we
consider coherent multiple Andreev reflection (MAR) regime, the theory of which has been mostly developed for the contacts with a single conducting channel: short junctions with the length $L\ll\xi_N$ \cite{MAR1995,MAR1995a,MAR1995b} and long junctions with a strong scatterer.\cite{MAR} In the multichannel transport regime in short junctions, the current can be evaluated by integrating the single-channel currents over the distribution of the normal junction transmission eigenvalues. The overall shape of IVC depends on the transmission distribution: the qualitative result for the dc current in short contacts is exponential decay at small voltage, $I\sim D^{2\Delta/eV}$ for various tunnel structures,\cite{MAR1995,MAR1995a,MAR1995b,BezuglyRC06,Our2011} while $I\sim V^{1/2}$ for diffusive constrictions.\cite{Averin1998} Furthermore, the IVCs contain current structures at the gap subharmonics, $eV=2\Delta/n$ (subharmonic gap structure).

In long multichannel junctions one would expect that the current resonances found in the single-channel junction \cite{MAR} will be averaged out after integration over channels, which is equivalent to averaging over their effective lengths. However, as we will show in this paper, the resonances survive, due to the length cut-off of the conducting channels, while the averaging results in a universal scaling behavior of the IVC in the limit of long junctions, accompanied by essential enhancement of the subgap current.

The structure of the paper is the following. After brief discussion of the MAR theory for a single channel and its generalization to the multichannel case in Section II, we present the results of numerical calculation of the IVC of a multichannel junction in Section III. This calculation indicates the existence of two qualitatively different regimes. In a relatively short junctions with only two Andreev levels for the majority of quasiparticles, the IVC has an exponential form and exhibits features at subharmonics of the maximum value $\delta_m$ of the interlevel distance. In the limit of a long junction (multilevel situation), the IVC is power-like, and the resonant features occur at the voltages which are multiples (harmonics) of $\delta_m$. In order to explain the physics of these results, we develop in Section IV an analytical approach to the calculation of the IVC for long multichannel junctions. We show that the single- and two-particle currents scale as $R_N^{-1}$, where $R_N\sim D^{-1}$ is the normal resistance of the junction, and the $n$-particle currents ($n>2$) are described by {\em universal} dependencies $I_n=(\Delta\sqrt{D}/eR_N) F_n(eV/\Delta)$, thus scaling as $D^{3/2}$.

\section{Calculation scheme}

In this Section, we briefly discuss our calculation method based on the results of calculation of the dc current for the voltage biased single-channel (one-dimensional, 1D) structure.\cite{MAR} In the 1D case, the net current $I^{\textrm{1D}}(V)$ is expressed through the sum of the contributions of partial $n$-particle spectral currents $j_n(E)$,
\begin{align}
I^{\textrm{1D}} &= -\frac{e}{h}\sum_{n=1}^\infty n \Bigl[
\Theta(neV - 2\Delta) \int^{-\Delta}_{\Delta - neV} dE\, j_n(E)
\tanh\frac{E}{2T} \nonumber
\\
&+ \int^{-\Delta-neV}_{-\infty} dE j_n(E)\Bigl( \tanh\frac{E}{2T}
-\tanh\frac{E_n}{2T}\Bigr)\Bigr],
\end{align}\label{j1D}
where $\Theta(x)$ is the Heaviside step function and
\begin{align} \label{jn}
j_n(E) &=\frac{8\xi\xi_n}{\Delta^2 |Z(E)|^2} \left(e^{\gamma} +
|r_{0-}|^2 e^{-\gamma}\right) \left(e^{\gamma_n} + |r_{n+}|^2
e^{-\gamma_n}\right),
\\ \nonumber
Z(E) &= \left(\begin{array}{ccc}1, & -r_{0-}\end{array} \right)
\left(\hat{U}_n\hat{M}_{n0}\hat{U}_0\right)^{-1} \left(\begin{array}{ccc} 1
\\ r_{n+}\end{array} \right).
\end{align}
Here the index $n$ indicates the shift of the energy, $E_n = E+neV$,
$\exp(\gamma) =(E+\xi)/\Delta$, and $\xi(E) = \sqrt{(E+i0)^2-\Delta^2}$ is the analytical function of the complex-valued energy defined in the upper half-space. The amplitudes $r_{n\pm}$ refer to the limiting values of the following ratios of matrix elements of the $2\times 2$ matrix $\hat{M}_{nm}$:
\begin{align} \label{r}
r_{n+} = - \lim_{m \rightarrow +\infty} \frac{\left(\hat{M}_{mn}\right)_{11}}{
\left(\hat{M}_{mn}\right)_{12}}, \quad r_{0-} = \lim_{m \rightarrow -\infty}
\frac{\left(\hat{M}_{0m}\right)_{12} }{ \left(\hat{M}_{0m}\right)_{22}}\,.
\end{align}
In practical calculations, the limits in \Eq{r} are truncated according to the rule: the energies $E_m$ and $E_n$ defined by the indices of the matrix $\hat M_{mn}$ are to be located at different sides of the energy gap, with few added steps in $m$ for better accuracy.

The matrix $\hat{M}_{nm}$ ($\det\hat{M}=1$, $n>m$) has the meaning of a transfer matrix along the MAR trajectory in the energy space across the gap,\cite{mapping} and is defined as the product
\begin{align} \label{M}
\hat{M}_{nm} &= \hat{T}_{n-1}\hat{U}_{n-1}\hat{T}_{n-2}\hat{U}_{n-2}\ldots
\hat{T}_{m},
\\
\hat{U}(E) &=  e^{-\sigma_z \gamma},\quad \hat{T}_{2n} =
\left(\hat{T}_{2n}^e\right)^{-1},\quad \hat{T}_{2n+1} =
\hat{T}_{2n+1}^h\,. \label{U}
\end{align}
Here the matrices $\hat{U}_n(E)=\hat{U}(E_n)$ describe the phase shifts of the wave functions associated with the Andreev reflection, and $\hat{T}_n^{e,h}=\hat{T}^{e,h}(E_n)$ are the real space transfer-matrices of the junction in the normal state for electrons and holes, respectively. We will consider a purely ballistic limit, neglecting electron scattering on random impurities; the potential barrier is modeled by a linear scatterer localized at the middle of the normal region and characterized with the energy independent scattering amplitudes, which will be assumed identical for all conducting channels. Furthermore, we consider transparent NS interfaces, which eliminates the Fabry-Perot interference effects and enables us to exclude the scattering phases, thus reducing the characteristics of the scatterer to its transparency $D$ and reflection coefficient $R=1-D$. Accordingly, the transfer-matrices $\hat{T}$ in \Eq{U} are composed with the transfer-matrix $\hat{t}$ of the scatterer, and the ballistic transfer-matrices $\hat{u}(E)$, which describe free propagation through the normal region,
\begin{align} \label{Tt}
\hat{T}^{e,h}(E) &= \hat{u}^{e,h}(E)\hat{t}\hat{u}^{e,h}(E), \quad
\hat{t} = D^{-1/2}(1 +\sigma_x \sqrt{R}),
\\
\nonumber \hat{u}^{e,h}(E) &= \exp\Bigl[i\Bigl(p_x L \mp\sigma_z
\frac{\varphi}{2}\Bigr)\Bigr], \quad  \varphi(E) = \frac{L}{
v_x}\Bigl(E+\frac{eV}{2}\Bigr)\,.
\end{align}
In planar junctions, $v_x$ and $p_x$ refer to the longitudinal (perpendicular to the barrier) components of the particle velocity and momentum, respectively. Then the matrix $\hat{T}_n$ can be written as
\begin{equation} \label{T}
\hat{T}_n = D^{-1/2} \bigl[\exp(i\sigma_z \varphi_n)- (-1)^n\sigma_x
\sqrt{R}\bigr].
\end{equation}
One can rewrite \Eqs{U}--\eqref{T} in a more convenient form, by combining the
matrices $\hat u^{e,h}$ with the $\hat U$-matrices rather than $\hat t$-matrices. In such representation,
\begin{equation} \label{Tnew}
\hat{T}_n = D^{-1/2} \bigl[1- (-1)^n\sigma_x \sqrt{R}\bigr], \quad
\hat{U}_n = e^{-\sigma_z(\gamma_n - iE_n L/v_x)}.
\end{equation}
Along with this transformation, we exclude the factors with unity moduli from $Z(E)$ in \Eq{jn}.

Proceeding with the summation over the channels, we assume their number ${\mathcal N}$ to be macroscopically large. Enumerating the channels by the values of their transversal momentum $p_{\perp}$, which is quantized in a multichannel 2D junction of width $w$ as $p_{\perp k} = \pi k \hbar/w$, $k = 0,1,\ldots$, we arrive to the following rule of the summation over the channels in the quasiclassical limit ${\mathcal N}=wp_F /\pi\hbar \gg 1$ where $p_F$ is the Fermi momentum,
\begin{align} \label{2D}
I^{\textrm{2D}} &=\sum_{p_{\perp k}<p_F} I^{\textrm{1D}}(v_{x,k})
\to \frac{w}{\pi\hbar}\int_0^{p_F} dp_\perp I^{\textrm{1D}}(v_x)
\\ &= {\mathcal N}\int_0^1 d\mu q(\mu)
I^{\textrm{1D}}(v_F\mu), \quad  \nonumber q(\mu) =
\frac{\mu}{\sqrt{1-\mu^2}}.
\end{align}
Here $\mu$ is the cosine of the angle $\theta$ of incidence of the quasiparticle at the NS boundary, as shown in Fig.~\ref{jofet}. For a 3D structure with the cross-section area $S$ and ${\mathcal N} = \pi S p_F^2/ (2\pi\hbar)^2$,
\begin{align} \label{3D}
I^{\textrm{3D}} &=\sum_{p_{\perp\, kl}<p_F} I^{\textrm{1D}}(v_{x,kl}) \to
\frac{\pi S}{ 2(\pi\hbar)^2}\int_0^{p_F}p_\perp dp_\perp I^{\textrm{1D}}(v_x)
\\
&= {\mathcal N}\int_0^1 d\mu q(\mu) I^{\textrm{1D}}(v_F\mu), \quad
\nonumber q(\mu) = 2\mu.
\end{align}
The quantity $\cal{N}$ can be then excluded through the relation to the normal conductance $R_N^{-1}=(2e^2/h)D\cal{N}$, which is a sum of contributions $(2e^2/h)D$ of separate conducting channels in the normal state.

In what follows, we assume the temperature to be much smaller than the energy
gap $2\Delta$ in the superconducting electrodes, which results in the following expression used in our calculations of the dc current,
\begin{align} \label{IRN}
I = \frac{1}{eR_N}\!\!\sum_{n \geq 2\Delta/eV}\int_0^1 \!\!d\mu q(\mu) J_n, \;\; J_n = \frac{n}{D} \int^{-\Delta}_{-neV/2}\!\!\!\!\!\!\! dE j_n(E).
\end{align}
In \Eq{IRN} we used the symmetry of the spectral density $j_n(E)$ in
\Eq{jn} with respect to the middle of the initial integration interval
$\Delta-neV < E < -\Delta$.

\section{Numerical results}

\begin{figure}[tb]
\centerline{\epsfxsize=8cm\epsffile{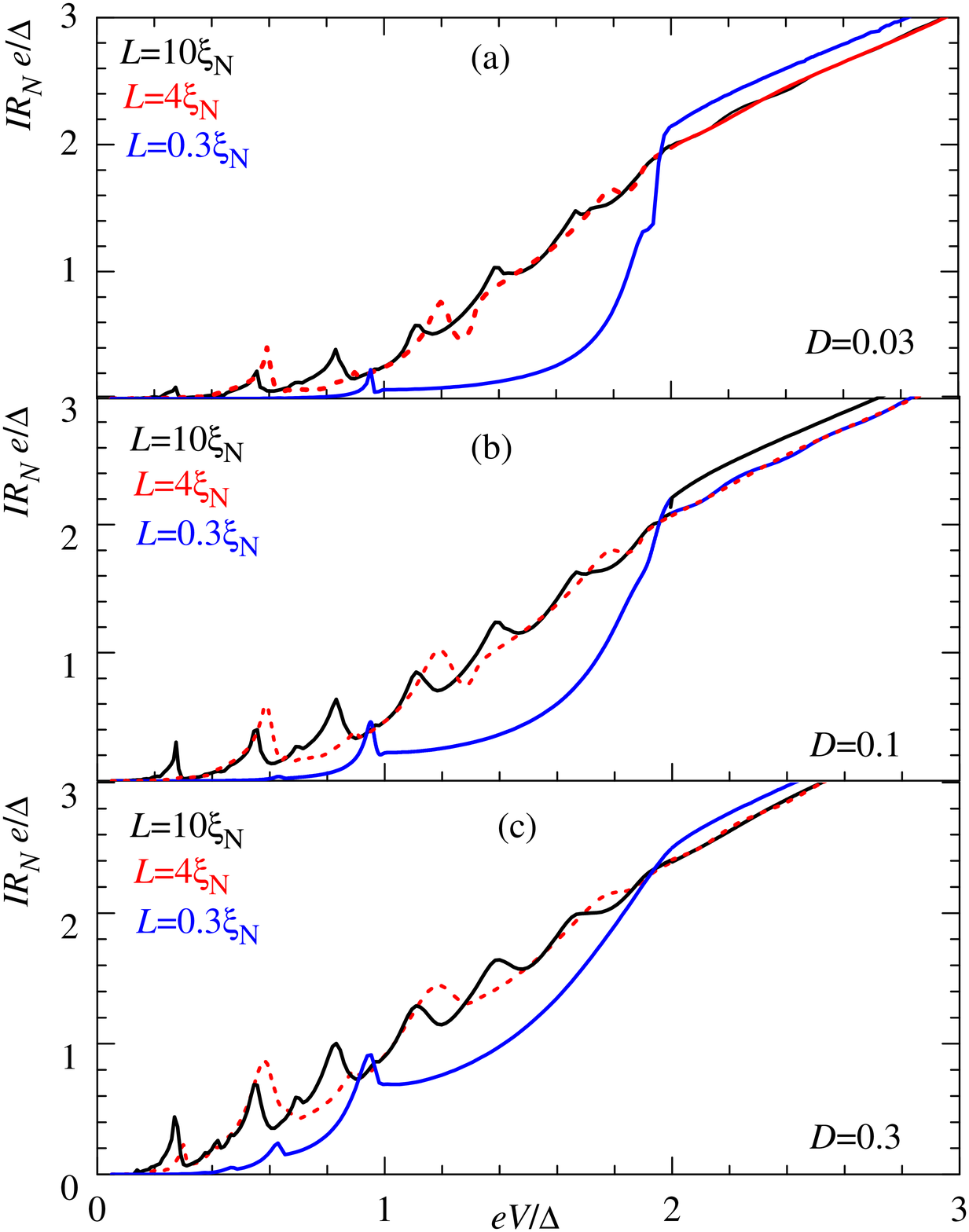}}\vspace{-4mm}
\caption{(Color online) Current-voltage characteristics for different transmission coefficients $D = 0.03$ (a), $0.1$ (b), $0.3$ (c) and lengths of the junction: $L/\xi_N=0.3$ (lower solid blue curves), 4 (red dashed curves), and 10 (upper solid black curves).} \vspace{-4mm}
\label{fullIVC2D}
\end{figure}

In Fig. \ref{fullIVC2D} we present the calculated IVCs for
three values of the transmission coefficients and lengths of the planar (2D)
SNINS junction (the results for 3D junctions are similar). For a short junction, $L=0.3\xi_N$, the current rapidly decays with decreasing voltage, similar to the case of a short single-channel contact; in the limit $L \to 0$, the results for all dimensions coincide (being normalized by $R_N$). On the contrary, for comparatively long junctions, $L = 4\xi_N$ and $L = 10\xi_N$, the current decreases with voltage much more slowly. Another interesting observation is that in the limit $L \gg \xi_N$, the smooth part of the IVC becomes independent of the junction length.

\begin{figure}[b] \vspace{-4mm}
\centerline{\epsfxsize=8.5cm\epsffile{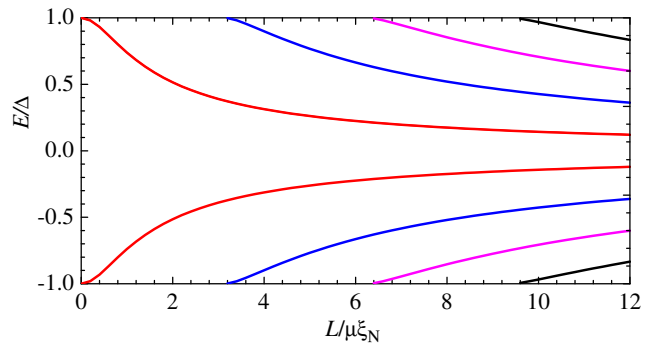}}\vspace{-4mm}
\caption{Emergence and evolution of Andreev levels with the increase of the
quasiparticle trajectory length.}
\label{Levels}
\end{figure}

The most of characteristic features of the IVCs in the junction of finite
length and width can be explained by the existence of discrete Andreev levels
in equilibrium junction, and corresponding singularities in the density of the quasiparticle states. It is instructive to qualitatively examine highly resistive junctions, where this connection is most prominent. In such junctions, the Andreev level spectrum can be approximated with the de Gennes-Saint-James levels \cite{dG} in each half of the normal layer with the length $L/2$ (the level splitting due to finite barrier transparency will be discussed later). Dispersion equation for these levels coincides with the one for a SNS junction of equivalent length $L$,
\begin{align} \label{AL}
& \frac{EL}{\Delta\mu\xi_N} = \pi n + \arccos\frac{E}{\Delta}, \;\; n=0,\pm
1,\pm 2, \dots,
\end{align}
where $|E|<\Delta$, $\mu = \cos\theta$, and $\theta$ is the angle between the
particle velocity and the normal to the junction interfaces. From \Eq{AL} we obtain simple estimates for the average interlevel distance $\delta$ and
the number of Andreev levels $n_A$,
\begin{align} \label{delta}
&\delta(L,\mu) \approx \frac{\Delta}{{L}/{\pi\mu\xi_N}+{1}/{2}}, \;\;\\
&n_A(L,\mu) = 2[\textrm{Int}({L}/{\pi\mu\xi_N}) +1],\label{nA}
\end{align}
[$\textrm{Int}(x)$ denotes integer part of $x$]. Evolution of Andreev levels
with the increase of the trajectory length $L/\mu$ is depicted in Fig.~\ref{Levels}.

\begin{figure}[tbh]
\centerline{\epsfxsize=8cm\epsffile{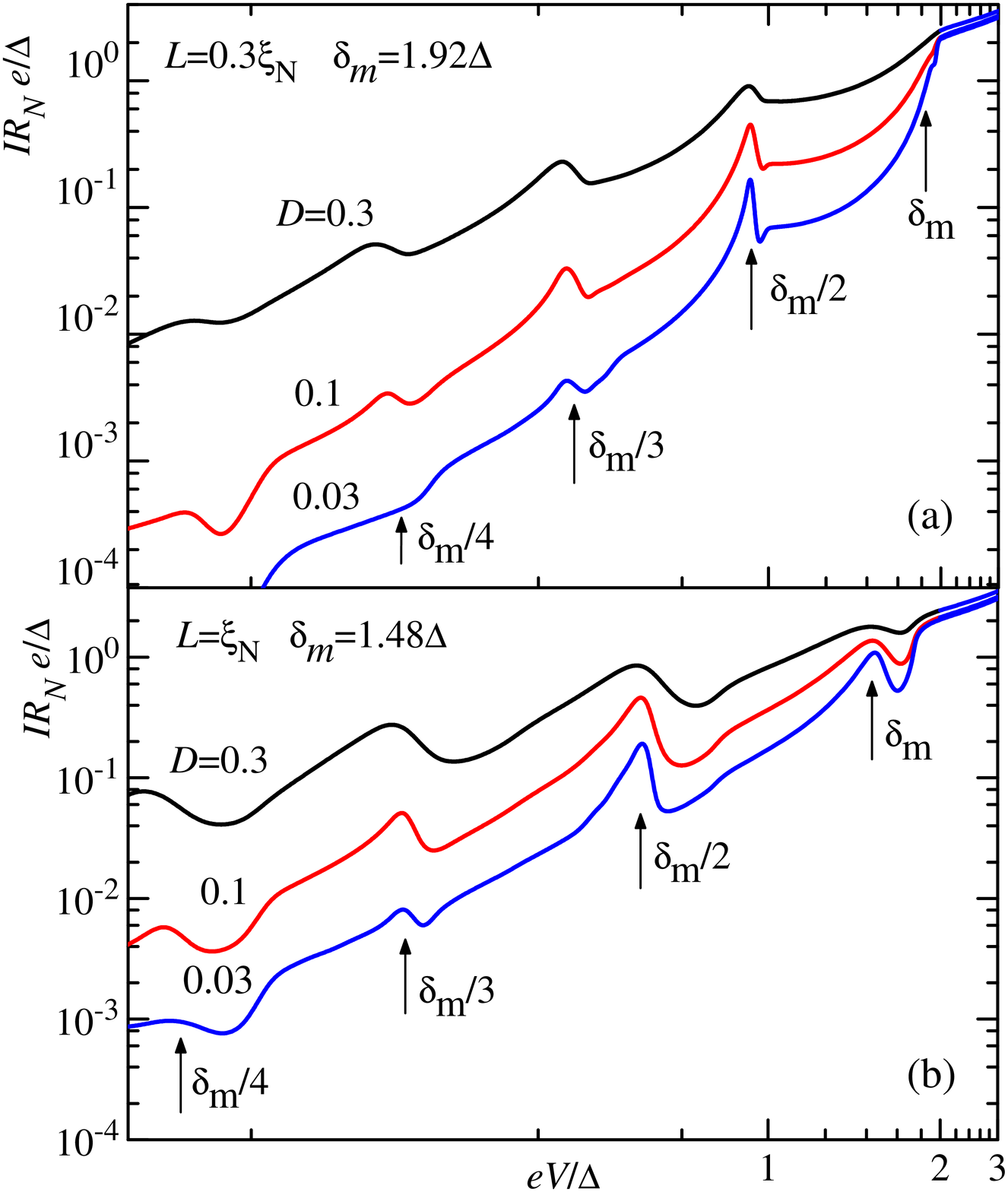}}\vspace{-4mm}
\caption{(Color online) Subharmonics $eV_n = \delta_m/n$ of the maximum interlevel distance $\delta_m$ for the junction lengths $L=0.3\xi_N$ (a) and $L=\xi_N$ (b) at different transmission coefficients.}\vspace{-4mm}
\label{Subharmonics2D}
\end{figure}
\begin{figure}[ht]
\centerline{\epsfxsize=8.3cm\epsffile{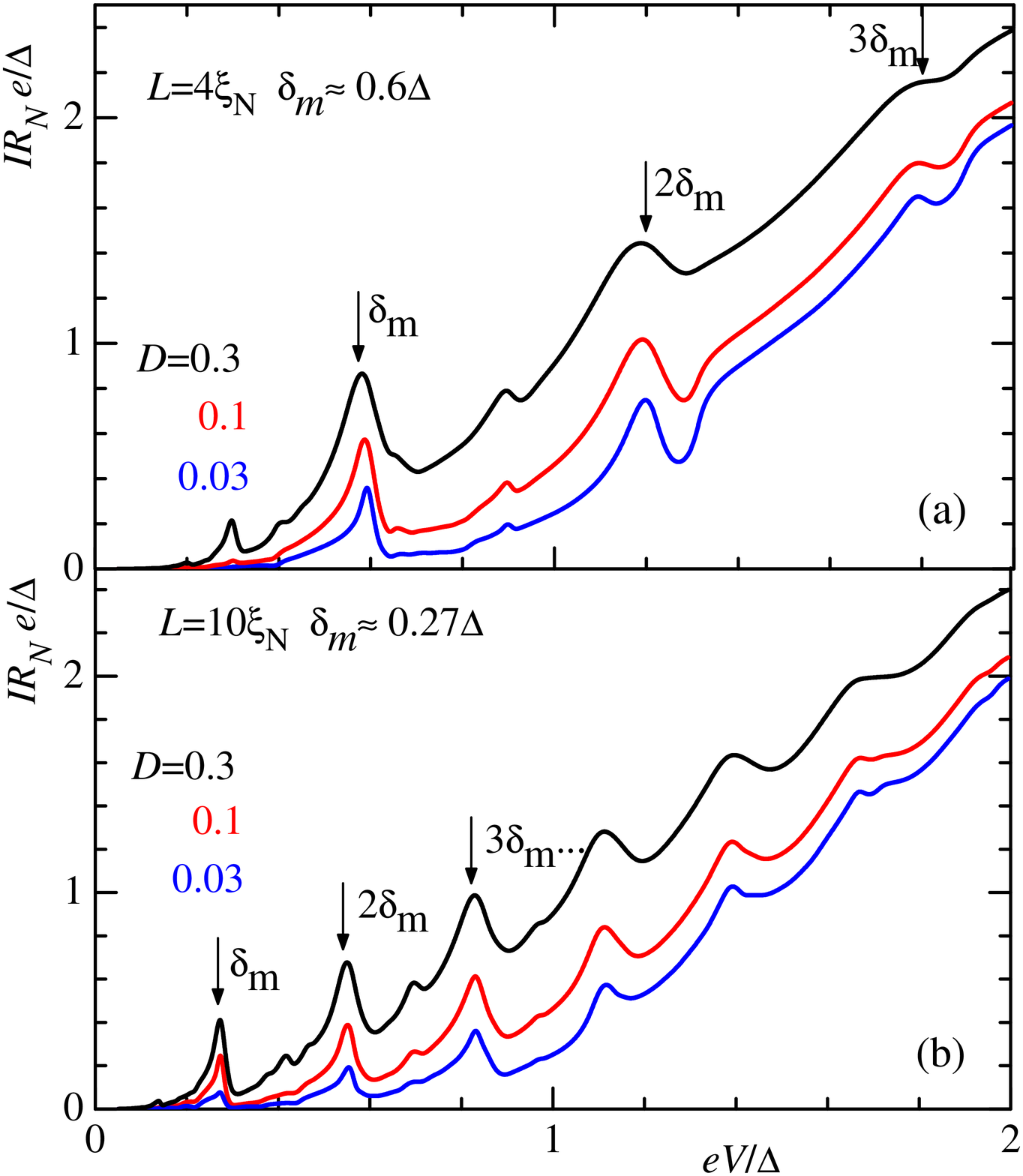}}\vspace{-4mm}
\caption{(Color online) Subgap features at multiple of the maximum interlevel distance $\delta_m$ for the junction lengths $L=4\xi_N$ (a) and $L=10\xi_N$ (b) at different transmission coefficients.}\vspace{-4mm}
\label{Subgap2D}
\end{figure}
\begin{figure}[ht]
\centerline{\epsfxsize=8.3cm\epsffile{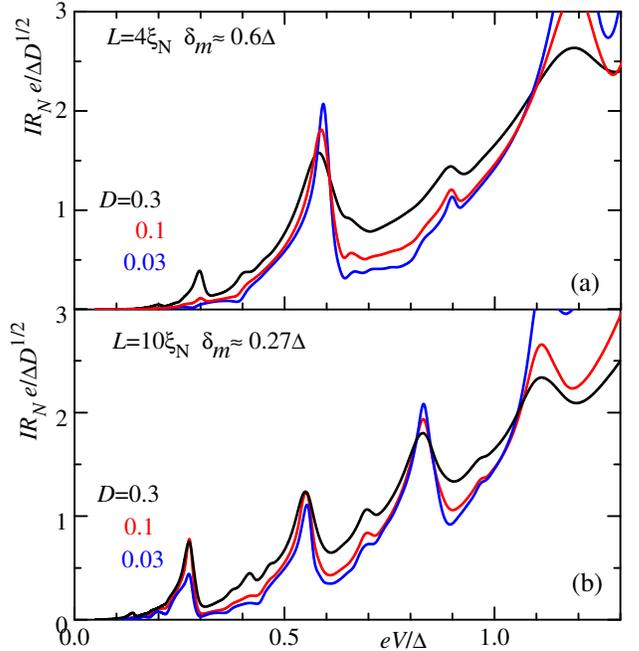}}\vspace{-4mm}
\caption{(Color online) Universality of the IVCs normalized over $\sqrt{D}/R_N$ at $eV< \Delta$ for the junction lengths $L=4\xi_N$ (a) and $L=10\xi_N$ (b) at
different transmission coefficients.}\vspace{-4mm}
\label{Universality}
\end{figure}

As follows from Fig.~\ref{Subharmonics2D}, the features on the IVC of relatively short junctions correspond to the subharmonics $eV=\delta_m/n$, $n = 1,2,\dots$, of the maximum distance between two Andreev levels present in the shortest electron trajectory, $\delta_m = \delta(L,1) < 2\Delta$, which is quantitatively different from common structure of the energy gap subharmonics. Obviously, these features are due to the resonant MAR trajectories in the energy space whose first and last Andreev reflections occur in the neighborhood of the Andreev levels where the density of states is singular. Similar effect has been found in the diffusive SNcNS junction\cite{Aminov} where the role of the potential barrier was played by the ballistic constriction (c), and the role of Andreev levels -- by the edges of the proximity induced minigap in diffusive normal metal banks with greatly enhanced density of states; appearance of the minigap subharmonics has been also predicted for a short diffusive double-barrier junction.\cite{Our2014} Furthermore, the averaged subgap IVC can be well approximated with the exponential dependence, $I(V) \sim D^{\const/eV}$, similar to the case of $L = 0$, but with the constant close to $\delta_m$ rather than to $2\Delta$.

A quite another picture of the IVC shape was found in lon\-ger junctions, where the number of Andreev levels is comparatively large: $n_A = 4$ for $L =
4\xi_N$, and $n_A = 8$ for $L = 10\xi_N$. In this case, the IVCs plotted in Fig.~\ref{Subgap2D} are non-exponential, and the resonant features correspond to multiples of the maximum interlevel distance, $eV = n\delta_m$, $n = 1,2,\dots$, or, in another words, to {\em harmonics} of $\delta_m$. From this we conclude that the enhanced multiparticle charge transfer is  entirely due to the resonant MAR trajectories for which all points of Andreev reflections in the energy space coincide with the level positions. We note that at very small voltages or transparencies, the nonequidistance of the Andreev levels plays role, and the fully resonant trajectories do not exist (for a more detailed analysis, see Ref.~\onlinecite{MAR}). This leads to a rapid decrease of the subgap current at small voltages clearly visible in Fig.~\ref{Subgap2D} at $eV \lesssim 0.3\Delta$.

The next intriguing feature of the IVCs for long junctions is their approximate overlap at $eV < \Delta$ for different $D$, provided $I(V)$ is normalized over $\sqrt{D}/R_N$, as shown in Fig.~\ref{Universality}. Note that the smaller $D$ is, the better this ``universality'' looks. This fact, as well as the universality of the averaged IVCs for large $L$ mentioned above, motivated us to develop the analytical theory which gives a clear physical explanation of the numerically found peculiarities in the IVCs.

\section{Analytical results }
\subsection{IVC at $eV \gg \Delta$: excess current }

We start with evaluation of the excess current, i.e., voltage-independent
deviation of the total current from the ohmic IVC at large applied voltage $eV
\gg \Delta$. At these voltages, an analytical result in the finite form can be obtained for all lengths and transparencies because only first two partial currents, $J_1$ and $J_2$, give non-vanishing contributions to the net current, and the amplitudes $r_{n\pm}$ approach the asymptotic values $r_{0-} = \sqrt{R}$, $r_{n+} = (-1)^n \sqrt{R}$. Within the first approximation, the current $J_1$ is given by $eV$, which results in a normal Ohm's law for the net current $I$. The contribution of $J_1$ to the excess current is related to the next term in its expansion over $\Delta/eV$ associated with the energies $E \sim \Delta$,
\begin{align} \nonumber
&J_1^{\text{exc}} = 2\int_\Delta^\infty dE\left\{
\frac{2\xi[2E-D(E-\xi)]}{ [2\xi + D(E-\xi)]^2 +4R\Delta^2\sin^2 \phi}
-\frac{E}{\xi}\right\},\\
&\phi(E,\mu) = E L/\Delta\xi_N \mu. \label{j1exc2}
\end{align}

\begin{figure}[t]
\centerline{\epsfxsize=8cm\epsffile{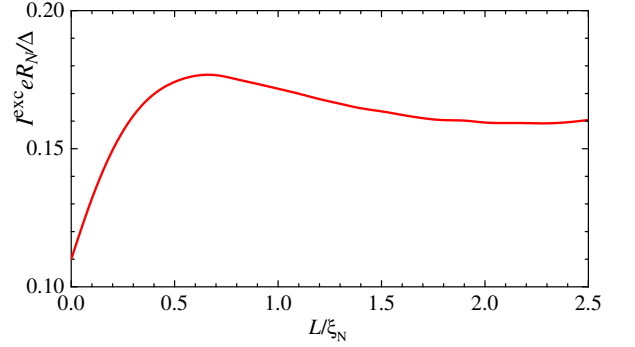}}\vspace{-4mm}
\caption{Excess current vs junction length for the
transparency $D=0.1$. The values of $I^\textrm{exc}$ in the limits $L \ll
\xi_N$ and $L \gg \xi_N$ are $0.11$ and $0.16$, respectively, in agreement with \Eqs{jexcL0} and \eqref{jexcosc}.} \vspace{-4mm}
\label{Excess}
\end{figure}

The contribution of the 2-particle current consists of two parts, in which the energy changes inside and outside the gap, respectively. The last contribution is combined with $J_1^{\text{exc}}$, and the general formula for the excess current reads
\begin{align}
&I^{\text{exc}} = \frac{2}{eR_N} \int_0^1 d\mu q(\mu) \left(2D\int_0^\Delta
\frac{dE}{D^2 +4R\sin^2 [\gamma(E)-\phi]}\right. \nonumber
\\
&\left. +\int_\Delta^\infty dE \left\{\frac{2E [ 2\xi+D(E-\xi)]}{
[2\xi+D(E-\xi)]^2 +4R\Delta^2\sin^2\phi}-\frac{E}{\xi} \right\}
\right), \label{jexcgen}
\end{align}
where $\gamma(E)=\arccos (E/\Delta)$. The behavior of the excess current as a function of the junction length is shown in Fig.~\ref{Excess}.

In the limit of a short junction, $L \ll \xi_N$, the main approximation in \Eq{jexcgen} reproduces known result for a single-mode junction (in
terms of the normal junction resistance),
\begin{align} \label{jexcL0}
&I^{\text{exc}} = \frac{\Delta}{eR_N} \frac{D}{R} \left[ 1- \frac{D^2}
{2\sqrt{R}(1+R)} \ln \frac{1+\sqrt{R}}{1-\sqrt{R}} \right]\!, \;\; L \ll \xi_N.
\\
\label{jexcL0app}
&I^{\text{exc}} \approx \frac{\Delta}{eR_N} \begin{cases}D, & D \ll 1; \\
8/3, & R \ll 1. \end{cases}
\end{align}
In the case of a long junction, $L \gg \xi_N$, the integrands in \Eq{jexcgen}
involve rapidly oscillating functions of $E$ and $\mu$. In order to obtain an
analytical result within the main approximation in $\xi_N/L$, we use the
following approach. Let us consider the integral
\begin{equation} \label{osc1}
A(\lambda)=\int_a^b dx f[x,\lambda z(x)],\quad \lambda \gg 1,
\quad f(x,y+2\pi) = f(x,y),
\end{equation}
where $f(x,y)$ and $z(x)$ smoothly vary at the distances much larger than
$\lambda^{-1}$. First, we introduce $z$ as a new variable and split the whole
interval of $z$ over small intervals of the length $p=2\pi/\lambda
\ll 1$,
\begin{align} \nonumber 
A(\lambda) &=\sum_{k=0}^{k_{\text{max}}-1} \int_{z(a)+kp}^{z(a)+(k+1)p} dz
f[x(z),\lambda z] \frac{dx(z)}{ dz}
\\ \nonumber
&+ \int_{z(a)+k_{\text{max}}p}^{z(b)} dz f[x(z),\lambda z]
\frac{dx(z)}{dz} , \quad k_{\text{max}}p < z(b)-z(a).
\end{align}

The second integral is of the order of $p\ll 1$ and can be neglected. In the
first term, we shift $z \rightarrow z+kp$ and use the periodicity of $f(x,y)$ with respect to $y$. Then, due to smoothness of the functions $f(x,y)$ and $z(x)$ with respect to $x$, we approximate the summation over $k$ by the integration over $z$ and then return to the integration over $x$ within the initial interval $(a,b)$, whereas the integral over the second variable reduces to averaging over the ``phase'' $u = \lambda z$,
\begin{equation} \label{osc4}
A(\lambda) \approx \int_a^b dx \int_0^{2\pi} \frac{du}{2\pi} f(x,u).
\end{equation}

Within such approximation, the first term in \Eq{jexcgen} is
\begin{equation} \label{jexcosc1}
2D \int_0^\Delta dE \int_0^{2\pi} \frac{du }{ 2\pi} \frac{1}{D^2 + 4R
\sin^2[\gamma(E) - u]}  = \frac{2\Delta}{1+R},
\end{equation}
whereas the second one is
\begin{align}
&\int_\Delta^\infty dE \int_0^{2\pi} \frac{du}{2\pi} \left\{\frac{2E [
2\xi+D(E-\xi)]}{[2\xi+D(E-\xi)]^2 +4R\Delta^2\sin^2 u}-\frac{E}{\xi} \right\}
\nonumber
\\
&= -\Delta\frac{1+R}{2R}\left(1-\frac{D}{\sqrt{R}}\arctan \sqrt{R}
\right),\label{jexcosc2}
\end{align}
which yields the final expression for the excess current,
\begin{align} \label{jexcosc}
&I^{\text{exc}} = \frac{\Delta}{eR_N} \frac{D}{R} \left( \frac{1+R}{\sqrt{R}}
\arctan\sqrt{R} - \frac{D}{1+R}\right), \;\; L \gg \xi_N,
\\ \label{jexcoscapp}
&I^{\text{exc}} \approx \frac{\Delta}{eR_N} \begin{cases} \pi D/2, & D \ll 1, \\
8/3, & R \ll 1. \end{cases}
\end{align}

By contrast to the case $L \ll \xi_N$, where both contributions to
the excess current are small ($\sim D$) at small transparencies $D$, the main
terms in \Eqs{jexcosc1} and \eqref{jexcosc2} are of the order of unity.
However, they have opposite signs and almost compensate each other,
which results in small value ($\sim D^2$) of the excess current in both limits. We note also that the expressions for $I^{\text{exc}}$ coincide for the short
and long junctions in the limit of high transparency, $R \ll 1$.

%

\subsection{Resonance approximation}

As we have seen from the numerical results, the subgap IVC of the short multichannel junction, $L \lesssim \xi_N$, is qualitatively similar to the IVC of a single-channel point contact, with the maximum interlevel distance $\delta_m$ replacing the superconducting energy gap $2\Delta$, and appropriately defined normal resistance $R_N$. A more interesting case, which allows us to obtain analytical results for arbitrary voltages, is the limit of a long junction, $L \gg \xi_N$, when the subgap charge transfer is enhanced due to existence of the resonant MAR trajectories which cross the energy gap through the chain of Andreev levels.

The single-particle current $I_1$ represents an exclusion: it does not involve Andreev levels and therefore always has a non-resonant nature. At $D \ll 1$ we have
\begin{equation} \label{I1}
I_1 = \frac{2}{eR_N} \Theta(eV-2\Delta) \int_0^1 d\mu q(\mu)
\int_{-eV/2}^{-\Delta} dE\, N\, N_1,
\end{equation}
where the function
\begin{equation} \label{DOS}
N(E,\phi)= \frac{E\xi }{\xi^2 + \Delta^2\sin^2 \phi(E)}
\end{equation}
has the meaning of the local density of states. At $L \gg \xi_N$, we apply the phase averaging method described in previous Section to perform integration over the energy and angle. This is done by averaging the integrand over two independent phases $\phi_0$ and $\phi_1$. This gives for $I_1(V)$ a linear      Ohmic dependence starting from the threshold point $eV=2\Delta$,
\begin{equation} \label{I1app}
I_1 = \frac{eV -2\Delta}{eR_N} \Theta(eV-2\Delta).
\end{equation}

Proceeding with multiparticle currents $I_n$, $n>1$, and taking into account only fully resonant trajectories, we must require the energies $E_k$ of all intermediate Andreev reflections ($1 \leqslant k \leqslant n-1$) to be located inside the gap, where the Andreev levels exist (actually, this means elimination of the MAR trajectories with partial over-the-barrier Andreev reflections above the energy gap). This results in the restriction $E < E_\textrm{max}$ for the initial energy of the MAR chain, and in a limited voltage range where the $n$-particle current is nonzero:
\begin{align}
I_n &= \frac{nD^{n-1}}{eR_N} \int^{E_{\text{max}}}_{-neV/2}
dE\,\left< \frac{8 N\, N_n }{|Z|^2}\right>,\label{In}
\\
E_{\text{max}} &= \text{min}[-\Delta,\Delta-(n-1)eV], \quad {2\Delta}/{n}
\leqslant eV\leqslant {2\Delta}/({n-2}), \nonumber
\end{align}
where the angle brackets denote averaging over two phases mentioned above.

The 2-particle current is a special one and differs from the high order currents: it involves only one intermediate Andreev reflection, and therefore only the initial energy $E$ should be adjusted in order to achieve the resonant transmission, when the energy $E_1 = E+eV$ coincides with the Andreev level; the interlevel distance is irrelevant in this aspect. Formally, this manifests itself in appearance of only one resonant phase in the denominator; a characteristic scale for this phase is $D$. The second independent phase is non-resonant and varies within the whole interval of periodicity,
\begin{align}
&I_2 = \frac{4}{eR_N} \Theta(eV-\Delta) \int_0^{2\pi} \frac{du}{2\pi}
\int_0^{\text{min}(\Delta,\,eV - \Delta)} \frac{dE}{N_+^{-1}+ N_-^{-1}} ,
\nonumber
\\
&N_\pm = N(E \pm eV, \phi_\pm), \quad \phi_\pm = \gamma(E) \pm u\label{I2}
\end{align}
[for symmetry, we used $E_1$ as the integration variable $E$ in
\Eq{I2}]. We note that due to resonant transmission through the intermediate
Andreev level, the 2-particle current appears to be of similar order $D$ as the single-particle one. This result was already discovered during the calculation
of the excess current; the latter, however, was found to be of the order of
$D^2$ due to cancelation of main terms in the 2-particle current and in the
``deficit'' part of the 1-particle current.

For high-order partial currents, $n>2$, both the energy of quasiparticles and the interlevel distance (the propagation angle) should satisfy the resonant conditions for given voltage, therefore two independent phases (other phases contain their linear combinations) are always resonant. Several results obtained with such an approach are presented in the Appendix.

\subsection{IVC at $eV<\Delta$}

The formal ``phase averaging'' method described above enables us to reproduce only the smooth part of the IVC; furthermore, the calculation of $I_n$ requires a separate approach for each $n$, and the complexity of expressions for $I_n$ enormously increases with $n$. In this Section, we formulate another semi-quantitative approach to the description of the IVC at $eV<\Delta$, which gives a more detailed and clear physical explanation of the resonant charge transmission. Within this approach we obtain simple and universal formulas showing a good agreement with the results of numerical calculations, including the resonant features of the IVC. This approach is based on the analysis for the single-channel contact\cite{MAR} which shows that the IVC exhibits a complex pattern of current peaks resulting from the multiple Andreev resonances. Integration over the angle will smoothen this pattern; however, one should expect that the reduced oscillations will remain since the geometric weight $q(\mu)$ of partial conducting channels has the threshold at the reference point $\theta=0$ ($\mu=1$). We will perform our estimations and analytical calculations within the model of equidistant Andreev spectrum, $\delta(E) = \const$, which formally corresponds to approximating of $\arccos(E/\Delta)$ in \Eq{AL} with a linear function. This model is justified by the numerical fact that for all transparencies (except of very small ones), the energy width of the multiparticle resonances is comparable with the non-equidistance of the Andreev levels.

\begin{figure}[tb]
\centerline{\epsfxsize=8cm\epsffile{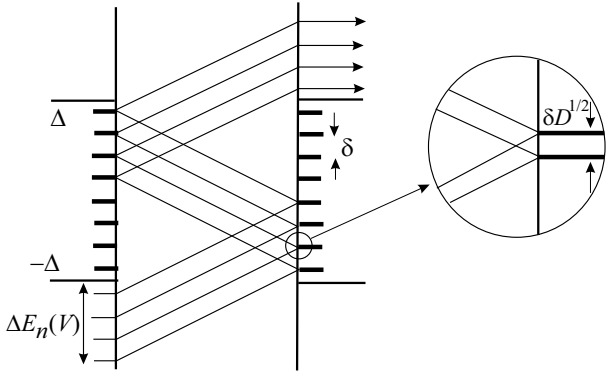}}\vspace{-2mm}
\caption{Resonant MAR trajectories passing through the Andreev levels (short bold lines) and contributing to the 3-particle current at $L=10\xi_N$ and $eV=4\delta$. The inset shows splitting of a static Andreev level to the cluster consisting of $2$ levels, therefore the whole number of the resonant trajectories is 8 in this case.}\vspace{-4mm}
\label{mar}
\end{figure}

As before, we will consider the resonant MAR paths when all $n-1$ Andreev reflections occur at the resonance energies within the superconducting gap, i.e., the voltage is assumed to be commensurate with the level spacing, $eV=k\delta$, and it belongs to the interval $V_{n}<V<V_{n-2}$ ($V_n=2\Delta/n$). Such trajectories, shown in Fig.~\ref{mar} for the 3-particle current, give the main contribution to the current spectral density $j_n \sim D(E,E+neV)$, where the quantity
\begin{equation}\label{Dmn}
D(E_m,E_n) = \left|\left(\hat{M}_{mn}\right)_{11} \right|^{-2}
\end{equation}
has the meaning of the effective transparency of MAR chains between the points
$E_n$ and $E_m > E_n$ at the energy axis.\cite{MAR} This resonant MAR process is mapped onto a periodic $n$-barrier tunnel structure with resonant levels in each well having the same energy when the barriers are non-transparent.\cite{mapping} At finite barrier transparency, the levels will repel each other forming a tight cluster of resonant levels within the energy interval $\sim \delta\sqrt D$ (precursor of the energy band at $n \to\infty$), as shown in the inset in Fig.~\ref{mar}. This leads to the splitting of the resonant transmissivity of the whole structure into $n-1$ peaks at the energy axis and results in the following schematic structure of the function
\begin{align} \label{jform}
D(E,E_n) = \frac{ D^n} {\prod_{k=1}^{n-1} (\epsilon/\delta -a_k\sqrt{D})^2 +
\Lambda^2}
\end{align}
in the vicinity of the cluster for a single conducting channel.\cite{MAR} Here $a_k$ are numerical constants describing distribution of the resonant levels over the cluster, and $\epsilon$ is the deviation of the MAR chain from the resonant position in the energy space. The maximum value and the width of the resonance in \Eq{jform} is determined by the magnitude of the second term in the denominator, which is of the order of $(n-1)^2 D^n$ at $n>2$.\cite{MAR} This results in the peak height $\sim 1/(n-1)^2$ and the width $(n-1)D\delta$, the latter being much smaller than the distance $\delta\sqrt{D}$ between the levels within the cluster. This means that all peaks in the resonant region $\delta\sqrt{D}$ give independent partial contributions into the net current, thus forming the total contribution of a resonant MAR chain,
\begin{equation} \label{jres}
J_n[\textit{chain}] \sim \frac{2e}{h}n  \sum_{j=1}^{n-1} \int_{-\infty}^\infty d\epsilon
\frac{D^2\delta^2}{\epsilon^2 + [(n-1)a_jD\delta]^2}\sim \frac{2e}{h} nD\delta.
\end{equation}

The number of such chains can be estimated as $\mathcal{E}_n(V)/\delta$, where $\mathcal{E}_n(V)$ is the width of the interval of the energy integration, in which the resonant trajectories exist (see Fig.~\ref{mar}, where four resonant sets of MAR trajectories are depicted). If the voltage $V<V_{n-1}$, then this interval is $\mathcal{E}_n(V) = neV-2\Delta = 2\Delta(V/V_n - 1)$ and increases with voltage; at larger voltage, $V>V_{n-1}$, the quantity $\mathcal{E}_n(V) = 2\Delta(1-V/V_{n-2})$ decreases and turns to zero at $V = V_{n-2}$. Thus, the resonant magnitude of the $n$-th current at $eV = k\delta$ is given
by
\begin{equation} \label{jpeak}
J_n^{res}\sim \frac{2e}{h} n D \mathcal{E}_n(V)
\end{equation}
and approaches maximum value $(2e/h)2n\Delta D/(n-1)$ at $V = V_{n-1}$. The
width of the resonance in voltage can be estimated as follows. The deviation in voltage, $\Gamma(eV)$, from its resonant value $eV = k\delta$ results in
deviation of the distance $(n-2)\Gamma(eV)$ between the first and the last
$n-1$ subgap Andreev reflections. In order to hold the resonant transmissivity, the latter value should not exceed the width of the resonant region
$\delta\sqrt{D}$. As the result, the width of the voltage peak at the
single-channel IVC \cite{MAR} is
\begin{equation} \label{Vwidth}
\Gamma(eV) \sim \delta\sqrt{D}/(n-2).
\end{equation}

In a multichannel junction, the resonant current magnitude in \Eq{jpeak} can be achieved at arbitrary subgap voltage by tuning the level spacing $\delta(\mu) = \mu\delta_{m}$ [a more precise estimate is given by \Eq{delta}]. Thus, the main contribution to $J_n$ comes from narrow vicinities of the resonant level spacings, $\delta_k =\mu_k \delta_{m} = eV/k$. The width $\Gamma(\mu)$ of the resonant regions of $\mu$ can be estimated similar to the width of the voltage peak in a single-channel case: the deviation in $\delta$, $\Gamma(\delta)= \Gamma(\mu)\delta_{m}$, times the number of interlevel distances $eV(n-2)/\delta$ between the first and the last subgap Andreev reflections, should not exceed $\delta\sqrt{D}$, which yields
\begin{equation} \label{muwidth}
\Gamma(\mu) \sim \frac{\delta^2 \sqrt{D}}{eV(n-2)\delta_{m}},
\end{equation}
and therefore the contribution from a given resonant value $\delta_k$ to the
current is
\begin{align} \label{Ik}
&J_n^{(k)}\sim {\cal N} j_n^{res} \left[ \Gamma(\mu) q(\mu)\right]_{\mu =
\mu_k}
\\
&\sim \frac{2e}{h}{\cal N} \frac{n D^{3/2} \mathcal{E}_n(V)}{ eV(n-2) \delta_{m} } \left(\frac{eV}{k}\right)^2 q\left(\frac{eV}{ k\delta_{m}} \right).\nonumber
\end{align}

Since $\delta_k$ can not exceed $\delta_{m}$, the contributions to the net
current $I_n$ come from $I_n^{(k)}$ with the numbers $k_{min}=
\textrm{Int}(eV/\delta_{m})+1 \leq k < \infty$. Thus, expressing the number of
channels ${\cal N}$ through the normal resistance, we obtain final expression for the $n$-particle current,
\begin{equation} \label{Itotsum}
I_n = C \frac{ n\mathcal{E}_n(V)\sqrt{D}}{ eR_N (n-2)\delta_{m}} \sum_{k =
k_{min}}^\infty \frac{eV}{k^2} q\left(\frac{eV }{ k\delta_{m}}\right), \quad n>2,
\end{equation}
where $C$ is a numerical constant. The value of this constant, $C=0.6$, is found by comparing \Eq{Itotsum} with the result of analytical calculation of multiparticle currents by using the ``phase averaging'' method in \Eq{In} (see Appendix).

Such approach can be easily generalized for the case of the angle-dependent transparency coefficient $D(\cos\theta)$: the constant factor $\sqrt{D}$ in \Eq{Itotsum} must be replaced by the function $\sqrt{D(eV/ k\delta_m)}$ within the sum over $k$. For a wide enough barrier, $D(\mu)$ rapidly decreases with $\mu$ and therefore plays the role of a cut-off factor for the contributions of very long trajectories associated with small resonant interlevel distances, $\delta_k = eV/k \ll \delta_m$ at large $k$. As the result, the sharp resonances at $eV = n\delta_m$ remain, but the smooth background will undergo noticeable suppression which brings the IVC closer to the one of the 1D case.\cite{MAR} Obviously, one should expect a qualitatively similar effect due to decreasing of the electron mean free path or the width $w$ of the junction.

At $eV \gg \delta_{m}$ ($k_{min} \gg 1$), we can approximately replace the
summation over $k$ by integration and obtain for both the 2D and 3D cases
\begin{equation} \label{triangle}
I_n(V) = C\frac{n\mathcal{E}_n(V)\sqrt{D}}{eR_N(n-2)}.
\end{equation}
In such approximation, the partial currents with numbers $n>2$, as functions
of voltage, have the shape of triangles with the bases between $V_n$ and
$V_{n-2}$ and with the apex at $V_{n-1}$ having the amplitude $2n\Delta
\sqrt{D}/ eR_N(n-1)(n-2)$. The net current $I(V)$ at given voltage $V$ ($V_n
< V < V_{n-1}$) contains contributions from two consequent partial currents: the increasing arm of $I_n$ and decreasing arm of $I_{n+1}$, therefore the IVC
represents piece-wise linear function connecting apexes of the triangles,
\begin{align} \label{I(V)}
&I(V) = \frac{C\sqrt{D} }{ R_N(n-2)}\left[ (n+2)V - \frac{4\Delta }{
e(n-1)}\right], \\ \nonumber
&V_n < V < V_{n-1}, \quad n >2.
\end{align}
Its behavior at the edges of of the subgap region is described by following
expressions,
\begin{equation} \label{Iasympt}
I(V)= C\frac{\sqrt{D}}{R_N}\begin{cases} 5V - 2\Delta/e, & eV
> 2\Delta/3, \\ V(1+eV/\Delta), & \delta_{m} \ll eV \ll \Delta. \end{cases}
\end{equation}
At smaller voltages, $eV < \delta_{m}$, \Eq{Itotsum} describes power-like decay of current with voltage, $I(V) \sim V^3$. Equations \eqref{I(V)} and \eqref{Iasympt}, as well as the more general
\Eq{Itotsum}, clearly demonstrate the universality of the averaged IVC found by the numerical computation: The smooth part of the dc current at $eV < \Delta$ is the universal function of the applied voltage, it is independent of the junction length, and scales as the $3/2$ power of the junction transparency.

\begin{figure}[th]
\centerline{\epsfxsize=8.5cm\epsffile{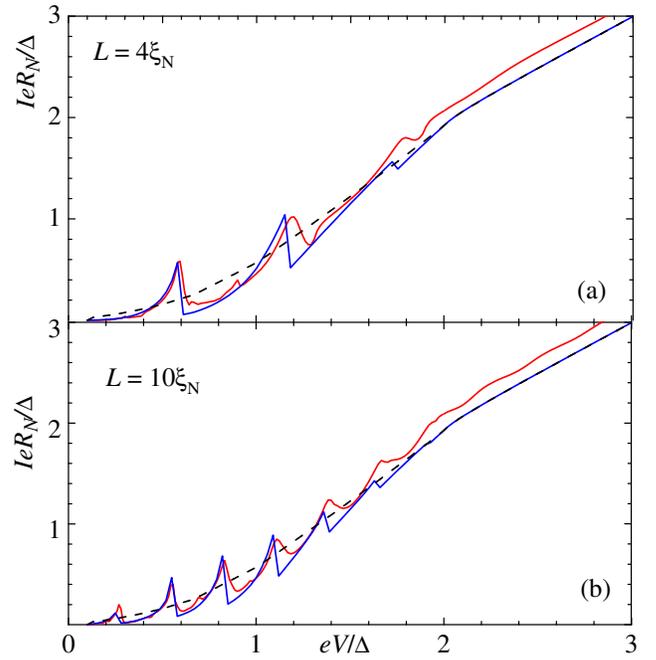}}\vspace{-4mm}
\caption{(Color online) Analytically calculated IVC [\Eq{Itotsum}, blue solid line] and its averaged form [\Eqs{I1app}, \eqref{I2} and \eqref{I(V)}, dashed curve], in comparison with the result of numerical computation (red solid curve) for a 2D junction with the transparency $D=0.1$ and length $L=4\xi_N$ (a), and $L=10\xi_N$ (b). } \label{Theory}\vspace{-2mm}
\end{figure}
\begin{figure}[th]
\centerline{\epsfxsize=8.5cm\epsffile{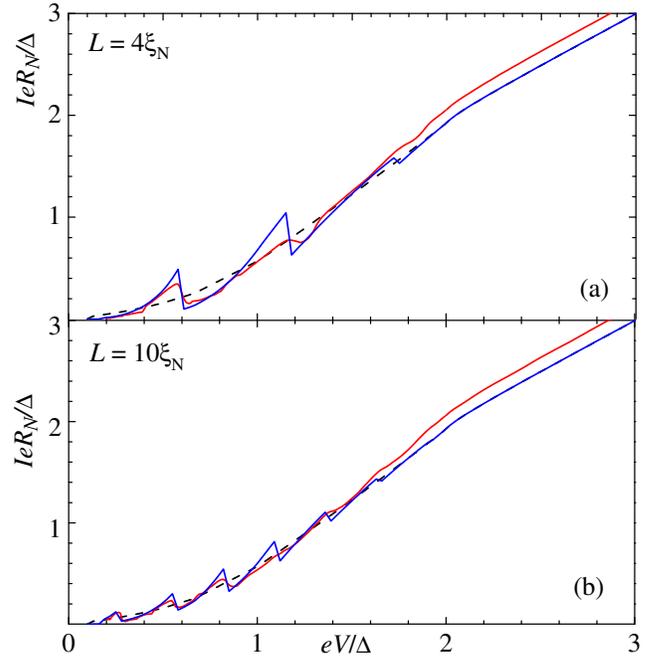}}\vspace{-4mm}
\caption{(Color online) The same as in Fig.~\ref{Theory} for the 3D junction. } \label{Theory3D}\vspace{-4mm}
\end{figure}

The comparison between the results of our numerical calculation and analytical
approach is shown in Fig.~\ref{Theory} where both the general \Eq{Itotsum}
(blue solid line) and and its averaged form \Eq{I(V)} (dashed line) were
plotted. We see that \Eq{Itotsum} also describes oscillations of the IVC
revealed in numerical computation with sharp peaks at the resonant voltages $eV = k\delta_{m}$ related to the electron trajectories at the reference point $\mu=1$. In the 2D case, due to divergence of the geometric weight $q(\mu)$ at the reference point, the formal calculation of IVC by means of \Eq{Itotsum} leads to singularities of $I(V)$ at the resonant voltages. A simple regularization of the geometric weight $q(\mu)$ by, e.g., a small shift of the singularity to the region $\mu>1$ eliminates these divergencies. The figure shows satisfactory agreement between the numerical and analytical results, except the absence of the excess current in the approximate expressions, \Eqs{I1app} and \eqref{I2}, for the 1-and 2-particle currents. Similar conclusion is valid for the 3D junction; in this case shown in Fig.~\ref{Theory3D}, no additional regularization is needed since the geometric weight is non-singular.

\section{Summary}

In this paper we investigate, both numerically and analytically, the dc current transport at given applied voltage through the ballistic SNINS junction of
finite length $L$ and width. In particular, such model corresponds to the
experimental setup with the normal 2D electron gas confined by massive
superconducting electrodes and controlled by a narrow electrostatic gate playing the role of a tunable tunnel barrier of moderately low transparency $D$ (Josephson field-effect transistor). Using the coherent multiple Andreev reflections formalism, we found that the characteristic features of the charge transfer in this device are fully determined by the presence of bound Andreev states which provide conditions for the resonant transmissivity.

In the limit of a short junction, $L<\xi_N$, when only one pair of the levels
exists for most of quasiparticles (except those propagating at small angles to
the interfaces), the IVC resembles the one of a ballistic point contact with a scatterer: the current decays exponentially with the voltage and exhibits a typical subharmonic structure periodic in $1/V$. However, in junctions with small but finite length, the role of the energy parameter, which defines the period of subharmonics and also the decrement of the exponential current decay, is played by the Andreev interlevel distance $\delta_m$ rather than the superconducting energy gap in the case of the point contact.

The most interesting phenomena were predicted for long junctions with multiple
Andreev levels. Existence of fully resonant MAR trajectories passing through
the chain of the levels essentially enhances the subgap current at voltages commensurate with the level spacing thus creating a resonance periodic structure in $V$ with the period $\delta_m/e$. Furthermore, the averaged IVC has a power form, and exhibits a peculiar universality: it does not depend on the junction length, and is universal for all junction transparencies at $eV < \Delta$ being normalized by $\sqrt{D}/R_N$. Physical explanation of these characteristic features, discovered in numerics, are given within the framework of the theory of resonant MAR charge transfer. We discuss the effects of the angle dependence of the transmission coefficient and the finite values of the electron scattering length and the contact width.

Finally, we note that both resonant effects -- subharmonics of the interlevel distance in short junctions and its harmonics in long ones -- can be used for direct experimental probing of Andreev levels in ballistic SNS structures.

\appendix*
\section{Multiparticle currents}

In this Section we present analytical expressions for several multiparticle
currents $I_n$ calculated in the limits $D \ll 1$, $L \gg \xi_N$ by making
use of \Eq{In} taken in the approximation of equidistant Andreev levels. Due to very cumbersome expressions for higher $n$-particle currents, we restrict our examples by $n=3$--$5$,
\begin{align} \label{I3}
&I_3 = \frac{3\sqrt{D}}{eR_N} \int^{E_{\text{max}}}_{-3eV/2} dE \sqrt{N_0
N_3},
\\
\nonumber &E_{\text{max}}=\text{min}\,(-\Delta,\Delta-2eV), \quad 2/3 <
eV/\Delta < 2,
\\ \nonumber &\phi_0 = 2\gamma_1 -\gamma_2, \quad \phi_3 = 2\gamma_2 - \gamma_1.
\\
&I_4 = \frac{2\sqrt{D}}{eR_N}  \int^{E_{\text{max}}}_{-2eV}
\frac{N_0 N_4 dE}{N_0+N_4} \biggl[\left(\frac{N_0}{N_4}\right)^{1/4}
+\left(\frac{N_4}{N_0}\right)^{1/4} \biggr],
\\
\nonumber &E_{\text{max}}=\text{min}\,(-\Delta,\Delta-3eV), \quad 1/2 <
eV/\Delta < 1,
\\
\nonumber &\phi_0 = (1/2)(3\gamma_1 -\gamma_3), \quad \phi_4 = (1/2)(3\gamma_3
- \gamma_1).
\\
\label{I5}
&I_5 = \frac{20\sqrt{2D}}{\pi eR_N}
\int^{E_{\text{max}}}_{-5eV/2} dE \left[\int_0^{y_-}+
\int_{y_+}^\infty\right]dy
\\ \nonumber
&\times \frac{N_0 N_5 N_+(y-2)\sqrt{y/R(y)}} {
(y-2)^2[R(y)+8y^2]N_+^2-R(y)(3y-2)^2N_-^2}
\\
\nonumber &R(y)=y^2-3y+1, \;\; y_\pm = (1/2)(3\pm\sqrt{5}), \;\; N_\pm = N_0
\pm N_5,
\\
\nonumber &E_{\text{max}}=\text{min}\,(-\Delta,\Delta-4eV), \quad 2/5 <
eV/\Delta < 2/3,
\\
\nonumber &\phi_0 = 3\gamma_2 -2\gamma_3, \qquad \phi_5 = 3\gamma_3 -
2\gamma_2.
\end{align}
Here $N_n= N(E+neV,\phi_n)$ is the density of states defined in \Eq{DOS}, and
$\gamma_k = \arccos[(E+keV)/\Delta]$.

\begin{figure}[th]
\centerline{\epsfxsize=8.5cm\epsffile{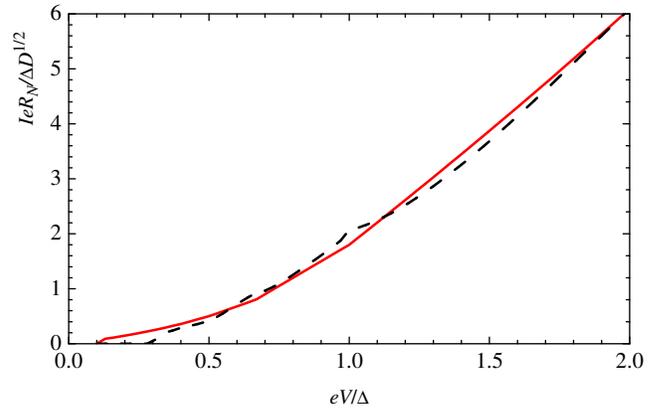}}\vspace{-4mm}
\caption{Subgap IVC calculated from \Eq{I(V)} with the fitting constant $C=0.6$ (solid line), compared with the sum of seven partial currents evaluated from \Eq{In} by the phase averaging method (dashed line).  }\vspace{-4mm}
\label{Fitting}
\end{figure}

These equations give the shape of $I(V)$ very close to \Eq{I(V)} which enables us to determine the fitting constant $C=0.6$ in \Eqs{I(V)} and \eqref{Itotsum}. The result of such fitting is shown in Fig.~\ref{Fitting}, where the dashed line is the sum of seven partial currents obtained in an analytically tractable form.

\end{document}